\documentstyle[epsf]{mn2e}

\DeclareMathAlphabet{\mathbi}{\encodingdefault}{\rmdefault}{\bfdefault}{\itdefault}
\DeclareRobustCommand{\bit}[1]{\ifmmode\mathbi{#1}\else\textbf{\textit{#1}}\fi}

\newcommand{\be}{\begin{equation}}
\newcommand{\ee}{\end{equation}}
   
\title[WMAP anomaly in the ecliptic plane.]
{WMAP anomalous signal in the ecliptic plane.} 

\author[Diego et al.]  
  {J.M Diego$^{1}$, M. Cruz$^{2}$, J. Gonz\'alez-Nuevo$^{3}$, M. Maris$^{4}$,Y. Ascasibar$^{5}$, C. Burigana$^{6}$\\  
   $^1$ IFCA, Instituto de F\'\i sica de Cantabria (UC-CSIC). Avda. Los Castros s/n. 39005 Santander, Spain.\\
   $^2$ Dept. de Matem\'aticas, Estad\'i stica y Computaci\'on. Universidad de Cantabria. Avda. Los Castros s/n. 39005 Santander, Spain.\\
   $^3$ SISSA-ISAS, via Beirut 4, I-34014 Trieste, Italy.\\
   $^4$ INAF - Osservatorio Astronomico Trieste, Via G.B.Tiepolo 11, I34100 Trieste, Italy.\\
   $^5$ Universidad Autónoma de Madrid, Departamento de Fisica Teorica, Madrid E-28049, Spain.\\
   $^6$ INAF-IASF Bologna, via Gobetti 101, I-40129 Bologna, Italy}
\date{Draft version \today}  
  
\pagerange{\pageref{firstpage}--\pageref{lastpage}}  
  
\begin{document}  
\maketitle  
  
\label{firstpage}  
\begin{abstract}  
We report the detection of a high Galactic latitude, large scale, 7-sigma signal in WMAP 5yr and spatially correlated with the ecliptic plane. Two possible candidates are studied, namely unresolved sources and Zodiacal light emission. We determine the strength of the Zodiacal light emission at WMAP frequencies and estimate the contribution from unresolved extragalactic sources. Neither the standard Zodiacal light emission nor the unresolved sources alone seem to be able to explain the observed signal. Other possible interpretations like Galactic foregrounds and diffuse Sunyaev-Zel'dovich effect also seem unlikely. We check if our findings could affect the low-l anomalies which have been reported in the WMAP data. Neither Zodiacal light emission nor unresolved point source residuals seem to affect significantly the quadrupole and octupole measurements. However, a signal with a quasi-blackbody spectrum and with a spatial distribution similar to the Zodiacal light emission, could explain both the anomalous signal and the low-$\ell$ anomalies. Future data ({\it Planck}) will be needed in order to explain the origin of this signal.
\end{abstract}  
\section{Introduction}\label{section_introduction}  
WMAP data (see Hinshaw et al. 2008 for a summary of the 5yr data release) revealed a wealth of information not only about the cosmological model but also about our own Galaxy. The most simplistic model describing the WMAP observations is comprised of the cosmic microwave background (or CMB hereafter), extragalactic point sources, Galactic components (synchrotron radiation, dust and free-free emission) all of them convolved with optical beams, and anisotropic instrumental noise. A combination of these 6 components can explain particularly well the observations at different channels over a relatively wide range of frequencies ($30$ GHz $< \nu < 90$ GHz). Some controversy arosed when anomalous residual signals appeared once the templates accounting for the components (with the exception of the noise for which no template can be produced) are subtracted from the original data (Finkbeiner 2004, Dobler \& Finkbeiner 2008). Detecting such residuals is not surprising (specially near the Galactic plane) since the spectral index is expected to vary across the sky for most of the components. These anomalous residual signals (or {\it anomalies}), could be the reason for the unusually low CMB quadrupole or for the alignment between octupole and quadrupole (another unusual feature), de Oliveira-Costa \& Tegmark 2006, Copi et al. 2006, Land \& Magueijo 2007). In Abramo et al. (2006) the authors study the existence of an hypothetical foreground which could explain the two mentioned anomalies. They showed how the anomalies could be explained as due to foreground residuals extending over an area of $\approx 0.5$ sterad, with a peak amplitude of about $20 \mu K$, a root-mean-square average of $8 \mu K$  and located in the ecliptic plane. Bunn \& Bourdon (2008) argue against this possibility that contaminants in the CMB can not explain the lack of power of the low-$\ell$ multipoles and that instead, this lack of power can be used as {\it a strong argument against the existence of undiagnosed foreground contamination}. The authors of this last work, though, consider that the possible contaminant is uncorrelated with the primordial CMB signal. This, of course, would not be the case of a signal which is aligned with the ecliptic and hence strongly correlated with the measured quadrupole and octupole in WMAP.

Here we present the detection of a new anomaly detected in WMAP 5-year data which appears at high Galactic latitudes and which resembles somehow the description given in Abramo et al. (2006).  This anomaly was first reported in Diego \& Ascasibar (2008) using WMAP 3-year data. We attempt to explain it as due to a combination of several components which have not been accounted for in previous works, the Zodiacal light emission (or ZLE hereafter) and the contribution from unresolved sources (US hereafter). The ZLE has been detected with a  high significance at higher frequencies by experiments like the Infrared Astronomical Satellite (IRAS) (Jones \& Rowan-Robinson 1993) or the Diffuse Infrared Background Experiment (DIRBE) (Kelsall et al. 1995). Extrapolations from these observations down to WMAP frequencies predict a weak signal, below the instrumental noise sensitivity at all scales (Maris et al. 2006). The predictions are made assuming that the dust has a single component and that the spectral frequency dependence of the intensity of the emission behaves as a blackbody with a temperature of $T\approx 240$ K modified by a factor $\lambda^{-2}$ at wavelengths $\lambda > 150 \mu m$ (Fixsen \& Dwek 2002). Multiple dust grain sizes are expected in the ecliptic plane so one would expect deviations from the $\lambda^{-2}$ law. In this paper we show how modifications of the  $\lambda^{-2}$ law can predict a signal which could be detected in the WMAP data. Analyzing WMAP data we find a signal which spatially correlates with the ZLE and with a dust-like spectrum. We also consider other possible signals which could produce a large scale signal like the one observed in WMAP. One of these possibilities is microwave emission due to unresolved sources (US). Finally we discuss the implications of this signal for aspects such as the measurements of the quadrupole and octupole in WMAP 5yr data.

\section{WMAP large scale residuals}
\begin{figure}  
   \epsfysize=4.5cm   
   \begin{minipage}{\epsfysize}\epsffile{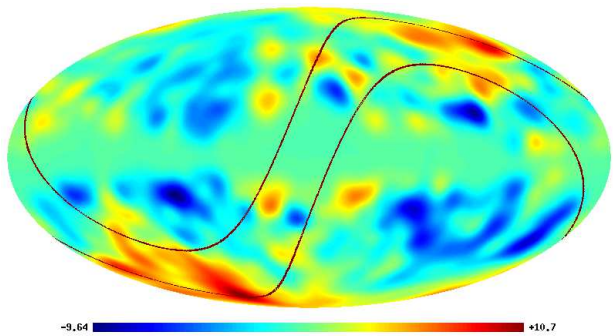}\end{minipage}  
   \caption{$7^{\circ}$ smoothed WMAP residual of the combination $V+W-2Q$. The color bar units are given in $\mu K$. The two curved lines show the ecliptic plane ($\pm 15^{\circ}$).
           }  
   \label{fig_WMAP}  
\end{figure} 
As a first (and good) approximation WMAP data can be decomposed into six components, namely CMB, resolved point sources, the three Galactic components (synchrotron, dust and free-free), and finally the leftover is assumed to be the contribution from instrumental noise. This simplistic decomposition manages to reproduce the observations with high accuracy. Nevertheless, one should expect other components to contribute as well to the WMAP observations. One of these components is the unresolved population of extragalactic sources. Only the brightest sources (a few hundreds) are masked out in the WMAP data. Many unresolved sources (US) still contribute to the data. This contribution can be particularly important if the US are correlated. For radio-selected extragalactic sources, their clustering signal introduces generally a small contribution to the temperature fluctuations thanks to the broadness of the local luminosity function (Dunlop \& Peacock 1990) and the redshift distribution of the sources, which dilutes the clustering signal (Blake \& Wall 2002, Gonzalez-Nuevo et al. 2005). In the case of the infrared sources, low-z galaxies dominate the counts at bright flux levels (Negrello et al., 2007) (ie. the kind of sources detected by the Infreared Astronomical Satellite IRAS, Beichman et al. 1998). Due to the inhomogeneous distribution in the local Universe, we expect overdensities in the flux distribution produced by the infrared sources. One can imagine that a diffuse component could be contributing to the data at scales similar to the correlation scale of the infrared sources. Such a signal would be easier to see once the small scales are filtered out. We use WMAP data to look for possible large scale residuals. Combining the different frequency channels and using Galactic foreground templates we can minimize the contribution from our Galaxy and the CMB itself. Filtering out the smaller scales, the instrumental noise contribution is reduced while the large scale residuals are highlighted. 
\begin{figure}  
   \epsfysize=12cm   
   \begin{minipage}{\epsfysize}\epsffile{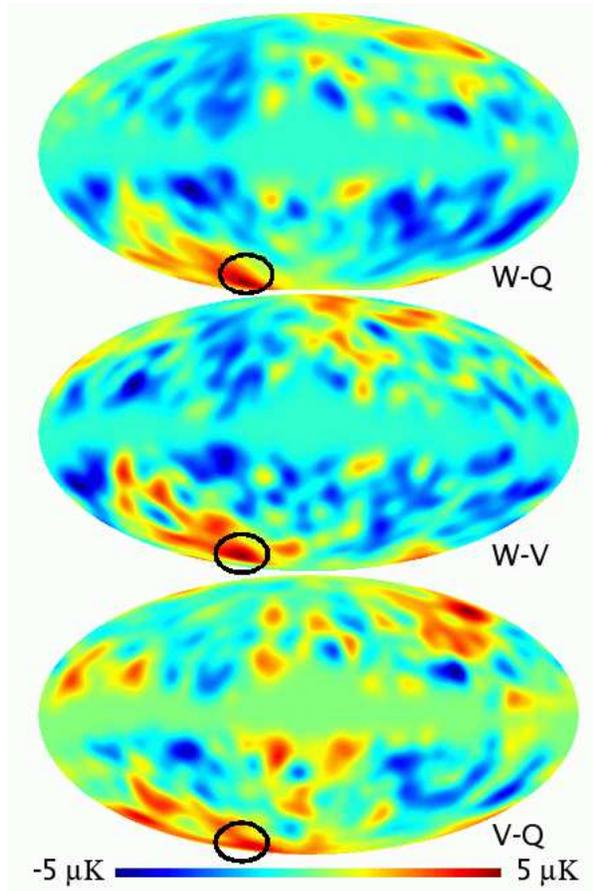}\end{minipage}  
   \caption{Differences after smoothing with a 7 degree Gaussian in the 3 individual 
            difference maps. The circle marks the spot where the amplitude 
            is measured in figure \ref{fig_Diffs2} 
           }  
   \label{fig_Diffs1}  
\end{figure} 
By residuals we mean components which are not considered in the cleaning process of WMAP 5yr data. The WMAP team provided foreground reduced maps where Galactic synchrotron, free-free and dust templates are subtracted from the data. We use these {\it clean}  maps and remove the CMB component by combining the $Q$, $V$ and $W$ frequency maps. In particular we choose the combination $V+W-2Q$. This CMB free combination should contain mainly noise. Signals with dust-like spectrum should appear as hot residuals while signals with a synchrotron-like spectrum should show up as cold residuals. There are other possible combinations which remove the primary CMB but we choose the mentioned one, following Diego \& Ascasibar (2008). Before combining the maps they are degraded to the resolution of the $Q$ band. We also subtract the monopole and dipole in the combined map to account for possible calibration differences between the channels. 

\begin{figure}  
   \epsfysize=6.0cm   
   \begin{minipage}{\epsfysize}\epsffile{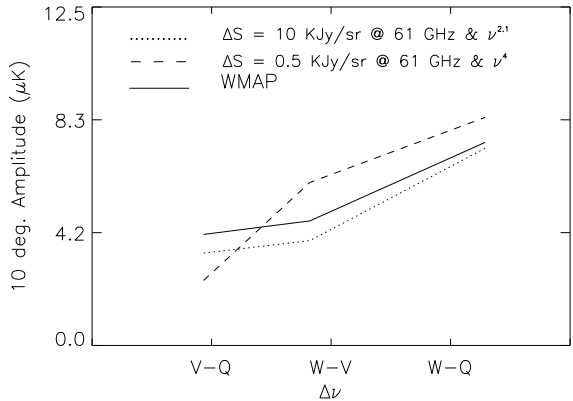}\end{minipage}  
   \caption{Amplitude at the positions marked in figure \ref{fig_Diffs1} 
            and for two different models. The solid line marks the amplitude 
            of the WMAP signal. The dotted line corresponds to a 
            model with an overdensity in flux of 10 KJy per steradian (at 61 GHz) 
            and with a emissivity law of $\nu^{2.1}$ (almost like a blackbody). 
            The dashed line is for a different model with an overflux 
            density of 0.5 KJy/sr and a emissivity law, $\nu^4$ (typical of thermal dust emission).    
           }  
   \label{fig_Diffs2}  
\end{figure} 

The final map is then smoothed with a 7 degree Gaussian kernel to reduce the instrumental noise contribution. The result is presented in figure \ref{fig_WMAP}. The map shows two very red regions (that is, with a dust-like spectrum) in the north-east and south-west quadrants. The central band with zero values is due to the exclusion mask used in the analysis. This mask also excludes the brightest point sources in WMAP data. On the other side, blue regions indicate areas with a possible synchrotron-like spectrum. The blue regions concentrate near the Galactic plane indicating that there might still be a residual Galactic synchrotron component in the combination map $V+W-2Q$.
This has been observed before (Bennett et al. 2003, Finkbeiner 2004). There is also a blue region near the Magellanic clouds indicating a possible contribution from spatially correlated radio US in this area. This paper will focus in the redder regions and will assume that the bluer regions are of Galactic origin. The Galactic interpretation for the redder regions is less likely due to their high latitudes. By looking at the individual differences (figure \ref{fig_Diffs1}) we notice that the redder regions appear in the three individual differences. In particular, the south-west redder region shows a clear pattern which is seen in the three differences. By looking at the intensity of the peak in this region (circles in figure \ref{fig_Diffs1}) we can infer the spectral index of the region. The WMAP intensity in the circled area is shown as a solid line in figure \ref{fig_Diffs2}. We compare this intensity with two models. The first one, represented by a dotted line, is a signal with an intensity of 10 KJy/sr in the V band and a emissivity law of $\nu^{2.1}$ (very similar to a blackbody law in the Rayleigh-Jeans region) in the V band. The second one, represented by the dashed line, is a signal with an emissivity law $\nu^4$ and intensity 0.5 KJy/sr (this model would correspond to thermal dust emission). The physical interpretation of the second model could be for instance a population of infrared sources with an overdensity of 0.5 KJy/sr in these regions. 
Since the redder regions correlate well with the ecliptic plane (which also has a modified blackbody spectrum), it is tempting to think that the redder zones could be due to ZLE from the ecliptic plane. It is also tempting to interpret the redder zones as due to a higher concentration of infrared extragalactic sources in these regions (a plot of the density of 2MASS extragalactic sources shows a clear overdensity of sources in these two regions, see figure \ref{fig_2MASS} below). But before we explore these two possibilities we should focus on another possible explanation. As the redder regions appear in areas of the sky where the instrumental noise is higher than the average, they could be just fluctuations of the noisy background in the combined map $V+W-2Q$. Whether the red regions are consistent or not with the instrumental noise is something we address in the next section.

\section{Statistical significance}
\begin{figure}  
   \epsfysize=7.5cm   
   \begin{minipage}{\epsfysize}\epsffile{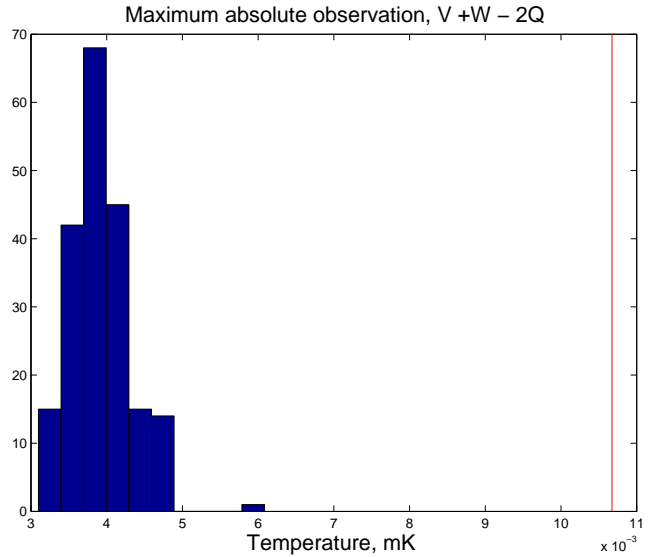}\end{minipage}  
   \caption{Histogram of the maximum of 200 simulated $V+W-2Q$ noise maps 
           (on a 7 degree scale). The vertical line between $10$ 
           and $12$ $\mu K$ is the actual observed value. 
           }  
   \label{fig_Histo1}  
\end{figure} 
To address the significance of the redder zones in the north and south hemispheres we perform 200 simulations of the instrumental noise. The noise in the $V+W-2Q$ map was simulated using the number of observations per pixel and per band given by the WMAP team and the instrument sensitivity. Independent noise maps were made for each one of the individual channels and those were combined in the way described in the previous section (i.e they are degraded to the resolution of the $Q$ map, combined and filtered with a Gaussian of FWHM $7^{\circ}$). We checked that $1/f$ noise is not an issue by using the noise simulations from the first year WMAP data (which include $1/f$ noise) and verified that no significant structures appeared on large scales due to $1/f$ noise. 
We look at the maxima of the combined simulated noise maps and compare them with the observed value of $\approx 10.7\times 10^{-3}$ mK. The result is shown in figure \ref{fig_Histo1}. We made histograms similar to the one shown in figure \ref{fig_Histo1} but for the individual differences ($W-V$, $W-Q$ and $V-Q$) and found similar results. The largest deviation was found in $W-Q$ (maximum of data at 7.5 $\mu K$ compared with maximum of simulations at 3.5 $\mu K$) followed by $V-Q$ (maximum of data at 4.5 $\mu K$ compared with maximum of simulations at 2.9 $\mu K$). The difference $W-V$ has a maximum in the data at 4.6 $\mu K$ which should be compared with a maximum in the simulations at 4 $\mu K$, (i.e, marginally consistent with the noise hypothesis).  
 The redder zones in the $V+W-2Q$ WMAP data are approximately a 7 sigma fluctuation with respect to the expected fluctuations of the instrumental noise. Hence, simulations show that the redder structures seen in the northern and southern hemispheres are not compatible with fluctuations of the instrumental noise and therefore they must be caused by a real signal. Some structures are visible in all the difference maps ($W-V$, $V-Q$ and $W-Q$). For instance, near the south Galactic pole and to the west, the red structure can be seen in all the difference maps (see figure \ref{fig_Diffs1}) suggesting that this is a real high latitude feature and neither a large scale fluctuation of the instrumental noise nor a systematic effect. In the next sections we will explore some of the possible explanations for the residuals. We will consider two possible components which combined with the instrumental noise might explain the observations. The first one is the ZLE, already detected at higher frequencies by DIRBE (Kelsall et al. 1995) and IRAS (Jones \& Rowan-Robinson 1993). The second one is the contribution from spatially correlated infrared sources. One extra possibility (the Sunyaev-Zel'dovich effect, Sunyaev \& Zel'dovich 1972) was already explored in a previous paper (Diego \& Ascasibar 2008) finding that the Sunyaev-Zel'dovich effect was an unlikely explanation. 

\section{Zodiacal light emission (ZLE)}
\begin{figure}  
   \epsfysize=4.5cm   
   \begin{minipage}{\epsfysize}\epsffile{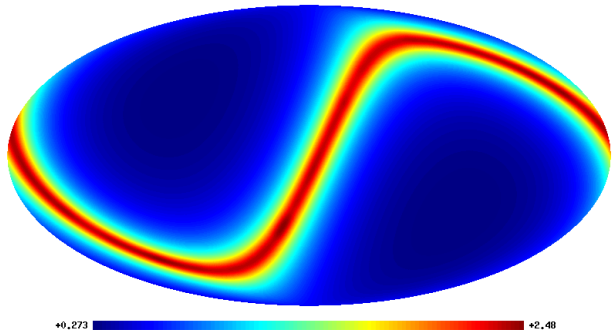}\end{minipage}  
   \caption{1 year average of the ZLE at 240 microns in Galactic coordinates. 
            The map has been filtered with a 7 degree Gaussian for presentation 
            purposes. The color table shows the units in MJy/sr. 
           }  
   \label{fig_Zodiacal}  
\end{figure}  
Dust grains in the ecliptic plane emit thermal radiation. Averaging this signal over one orbital year we can see a pattern which follows the ecliptic plane. This pattern is  shown in figure \ref{fig_Zodiacal}. This radiation is known as the Zodiacal light emission (ZLE) and it is mainly seen in the infrared, although it can also be seen in the optical since the dust particles scatter the light of the sun. The figure shows how the plane with the maxima of ZLE crosses the redder regions in the WMAP data. At sub-mm frequencies the ZLE was observed by DIRBE and IRAS. For our purposes we will follow Kelsall et al. (1998) to perform the simulation of the ZLE. The simulation was made at 240 microns using the code written by B.A Franz which implements the model described in Kelsall et al. (1998). In that work, the authors assumed multiple components for the dust density distribution: a smooth cloud, three asteroidal dust bands, and a circumsolar ring near 1 AU. This model manages to describe well the ZLE observed by DIRBE from from 1.25 to 240 microns. We extrapolated this signal down to WMAP frequencies using the following law:

\begin{equation}
Z_i  = F_i Z_{240}
\end{equation}
where $Z_i$ is the intensity of the ZLE in the WMAP band $i$ 
($i$ stands for the $\nu=41$ GHz or Q band, $\nu=61$ GHz or V band and $\nu=94$ GHz or 
W band), and $F_i$ accounts for the frequency dependence of the ZLE.
$Z_{240}$ is the template of ZLE as predicted by the code described above at 
240 microns (and given in $MJy/sr$ units). The ZLE map at 240 microns was pixelized using {\small HEALPIX} (Gorski et al. 2005) and transformed into Galactic coordinates. For the frequency dependence we assumed the following law.  
\begin{equation}
F_i =  g_i \left( \frac{B_i}{B_{240}}  \right)\left( \frac{240 \mu m}{\lambda _i(\mu m)} \right)^{\alpha} ~\mu K ~MJy^{-1} ~sr
\end{equation}
where $\lambda _i = 3 \times 10^5/\nu_i(GHz)$ in microns ($\nu_i$ is either 41 ($Q$), 61 ($V$) or 94 ($W$) GHz) and $B_i$ is the blackbody emissivity law and for $T=240 K$ (with $B_{240}$ the corresponding emissivity at 240 $\mu m$). The factor $g_i$ accounts for the conversion from $MJy/sr$ to $\mu K$ ($g_Q = 20.2 \times 10^3$, $g_V = 9.62 \times 10^3$ and $g_W = 4.6 \times 10^3$ all in $\mu K (MJy/sr)^{-1}$ units. When combining the different frequency maps, a combined $F$ factor can be defined as follows,
\begin{equation}
F_{comb} = F_V + F_W - 2F_Q ~ {\rm \mu K ~MJy^{-1} ~sr}
\end{equation}
Hence, the intensity of the ZLE (in $\mu K$ units) in the combination $V+W-2Q$ can be easily computed from the template $Z_{240}$ (shown in figure \ref{fig_Zodiacal}) as 
\begin{equation}
Z_{comb} = F_{comb} \times Z_{240} ~\mu K
\label{eq_Z}
\end{equation}

\begin{figure}  
   \epsfysize=6.0cm   
   \begin{minipage}{\epsfysize}\epsffile{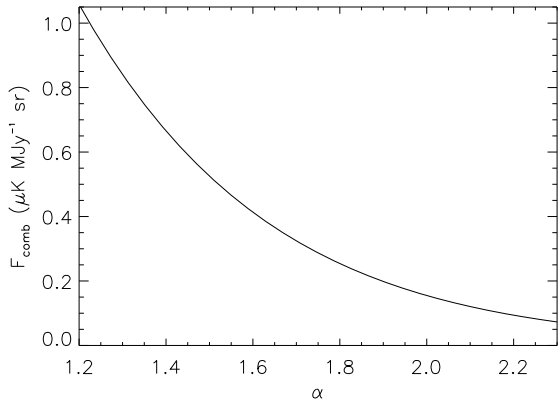}\end{minipage}  
   \caption{Dependency of the conversion factor ($F_{comb}$) with the exponent $\alpha$.
           }  
   \label{fig_Falpha}  
\end{figure}  

The ZLE light behaves like a greybody at wavelengths shorter than 150 $\mu m$ but at longer wavelengths it falls off more quickly. The fall off can be computed as a $\nu^{\alpha}$ correction to the blackbody spectrum. In  Fixsen \& Dwek 2002 $\alpha$ was found to be around 2 in the submillimeter to millimeter region, but the error bars allow a range of $\alpha$'s specially at longer wavelengths. A value of $\alpha=2$ corresponds to emission dominated by a population of dust grains with a narrow range of sizes. A deviation from the $\alpha=2$ exponent is expected when the ecliptic plane is filled with grains of different sizes. The value of $\alpha$ affects the amplitude of the predicted ZLE signal. We show this in figure \ref{fig_Falpha} where the factor $F_{comb}$ is computed for different values of $\alpha$. From figure 2 in Fixsen \& Dwek (2002) a valid range for $\alpha$ would be  $1.5 < \alpha < 2.5$ for wavelengths shorter than 1000 $\mu m$. It is unclear how the ZLE light behaves at $\lambda > 1000 \mu m$ and one could possibly think of effective $\alpha$'s smaller than 1.5 (see figure \ref{fig_Falpha}). In this work we will consider the case (consistent with the results of Fixsen \& Dweek 2000) where $\alpha = 1.5$, bearing in mind that smaller $\alpha$'s at WMAP frequencies can not be ruled out by current data. Extrapolating the predicted ZLE intensity from 240 microns to the WMAP frequencies with the value of $\alpha$ assumed above, we predict an intensity for the Zodiacal signal in the combined $V+W-2Q$ which is significantly smaller than the intensity of the redder zones in the north and south hemispheres in the $V+W-2Q$ difference maps. We predict a maximum amplitude for the ZLE of $\sim$ 1 $\mu K$ in $V+W-2Q$ after filtering with a 7 degree Gaussian which should be compared with the $\sim 10$ $\mu K$ amplitude observed in WMAP data. This can also be seen by combining figures \ref{fig_Zodiacal} and \ref{fig_Falpha}. In figure \ref{fig_Zodiacal} the maximum amplitude is 2.4 MJy/sr at 240 $\mu m$ which multiplied for the corresponding $F_{comb} \approx 0.5$ ($\alpha = 1.5$ in figure \ref{fig_Falpha}) renders a maximum amplitude of $\sim 1 \mu K$. From figure \ref{fig_Falpha} we see that if $\alpha \sim 1$ the contribution from the ZLE becomes more important, but is still far from the required $10 \mu K$. Hence we can conclude that it seems unlikely that the redder zones are due to standard ZLE emission.  

\section{Unresolved sources}
Another possible source of contamination in the WMAP data is the emission from unresolved extragalactic (or even compact Galactic) sources. In general terms, we will refer to these sources as unresolved sources (US). WMAP masks out the brightest sources but this removes only a few hundred sources above 0.5 Jy. Many other bright sources remain in the data. 

\begin{figure}  
   \epsfysize=4.0cm   
   \begin{minipage}{\epsfysize}\epsffile{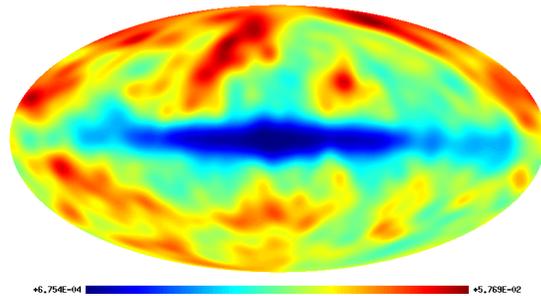}\end{minipage}  
   \caption{2MASS map of the number density (galaxies per $3.9 \times 3.9$ 
            arcmin$^2$ pixel) of extragalactic sources smoothed with a 10 degree 
            Gaussian. The lack of extended sources in the galactic plane 
            is due to the high density of foreground stars from our Galaxy.
           }  
   \label{fig_2MASS}  
\end{figure}

\begin{figure}  
   \epsfysize=4.0cm   
   \begin{minipage}{\epsfysize}\epsffile{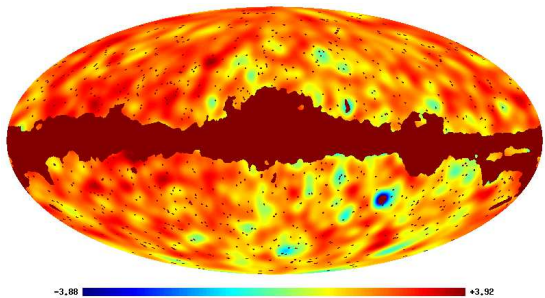}\end{minipage}  
   \caption{Predicted contribution from US with fluxes below 0.5 Jy in $V+W-2Q$ and 
            in $\mu K$. The WMAP mask is shown as well. Note the negative 
            contribution around the Magellanic cloud.
           }  
   \label{fig_PS}  
\end{figure}

On small scales (arcminute), the signal coming from these sources can be comparable to the CMB in terms of power spectrum amplitude. On larger scales (degrees) the amplitude of the power spectrum due to US is negligible compared to the CMB power spectrum. However, a combination map like the one used in this work ($V+W-2Q$) removes the primary CMB background leaving the US as a possible significant contributor to the residuals (specially if the US are clustered). The contribution coming from extragalactic sources might increase their integrated signal in certain regions over scales of many degrees if these sources are clustered. The most significant contributions are expected to come from nearby extragalactic clustered sources, since the flux of a source is inversely proportional to the square of its distance to us. In addition, at larger distances the clustering scales of the sources are under a few degrees (where the CMB clearly dominates) and the distribution of sources is also more homogeneous over the sky. The IRAS all sky map, for instance, is a good tracer for the possible contribution of nearby clustered infrared sources. Another example is the 2MASS where one can see clearly the inhomogeneous distribution of flux coming primarily from the local Universe (figure \ref{fig_2MASS}). We perform an all-sky simulation of the US under 0.5 Jy using the Green Bank 4.85 GHz northern sky survey (GB6, Gregory et al. 1996) and the Parkes-MIT-NRAO at 4.85 GHz (PMN, Gregory et al. 1994) catalogs to simulate radio sources at WMAP frequencies. The {\it holes} of these two surveys are filled with the NRAO VLA Sky Survey at 1.4 GHz (NVSS, Condon et al 1998) and the Sydney University Molonglo Sky Survey at 0.84 GHz (SUMSS, Mauch et al. 2003). For the infrared side we use data from the Infrared Astronomical Satellite (IRAS, Beichman et al. 1998) which is extrapolated down to WMAP frequencies.
In figure \ref{fig_PS} we show the predicted contribution from US in the $V+W-2Q$ map including only sources with fluxes below 0.5 Jy at each frequency ($Q$, $V$ or $W$). The US contribution is filtered with a 7 degree Gaussian and the result shows fluctuations of the order of a few $\mu K$ in $V+W-2Q$, but still far away from the observed 10 $\mu K$ level. Given our model, the combination of ZLE, US and instrumental noise could roughly explain half the amplitude of the observed signal in WMAP data. 

\section{Other candidates. Galactic emission and Sunyaev-Zel'dovich effect}

Whatever the cause of the redder zones is, we know that it must have a $\nu^{\alpha}$ spectrum with $\alpha > 2$. There are two other candidates, namely thermal dust emission from the Galaxy and the Sunyaev-Zel'dovich effect. The possibility that the redder zones are due to thermal dust in our own Galaxy is unlikely since the two red regions are close to the Galactic poles where the dust emission is minimal. Nevertheless, the redder zones could still be explained if the cleaning process removed more dust than needed, leaving the northern and southern Galactic poles with a positive {\it relative signal} when computing the individual differences of the WMAP channels. We consider this possibility by looking at the individual differences in the Galactic plane (without the mask) where one should expect a negative temperature difference and instead a small positive difference is observed (indicating that the dust template has not been over-subtracted). The other possibility is a diffuse Sunyaev-Zel'dovich effect coming from massive distant structures. Even though the Sunyaev-Zel'dovich effect produces a decrement in the temperature maps, when computing the differences $V-Q$, $W-V$ and $W-Q$, it produces a positive signal in all three differences. It is possible to constrain the expected Sunyaev-Zel'dovich by using X-ray measurements. In X-rays, the most luminous extragalactic object is the Virgo cluster. This cluster is also expected to be the brightest Sunyaev-Zel'dovich effect in the sky in terms of integrated flux (Diego \& Ascasibar 2008). In a previous work (Diego \& Ascasibar 2008), the authors showed how the Virgo region (which happens to be near the redder zone in the north) can contribute with a maximum of about 5 $\mu K$ at the center of Virgo when filtered with a 7 degree Gaussian. This prediction was made from models consistent with X-ray observations in the Virgo region. The amplitude falls quickly below $5 \mu K$ at angular distances of a few degrees from the center of Virgo. In the same work, the authors also estimated the possible contribution from a very extended (9 Mpc) Warm-Hot Intergalactic Medium (WHIM) cloud around Virgo, also consistent with X-ray constraints and concluded that the hypothetical WHIM cloud would contribute with about an extra $1 \mu K$. We can then conclude that the Virgo region could explain part of the observed signal in the north red zone but not the one located in the south. Another possibility is a very massive and distant structure with a hot and extended gas component. Simple calculations based on the total flux (see Diego et al. 2002) predict that models explaining the red zone in the Virgo region would exceed the X-ray constraints (in the same region) imposed by ROSAT observations. This model assumes a standard $\beta$ profile for the gas, temperature of 3 keV (isothermal), the structure is located at redshift 0.2 and the size of the structure is such that its apparent angular size is about 5 degrees. the mass of such a structure should be $\approx 1.5 \times 10^{15} h^{-1} M_{\odot}$. Other gas distributions with a different profile could potentially be consistent with the X-ray constraints but there is no evidence that such distributions could exist. It seems clear that the Sunyaev-Zel'dovich explanation is unlikely for the redder zones and we shall conclude that these regions still remain unexplained. 

\section{Implications for the measured quadrupole and octupole of the CMB}
If a combination of the ZLE and extragalactic US are affecting the WMAP data on large scales, it is possible that it will affect the measurements of the low multipoles. This possibility is even more realistic if we realize that the low multipoles seem to be aligned in a direction which follows the ecliptic (de Oliveira-Costa \& Tegmark 2006, Copi et al. 2006, Land \& Magueijo 2007). So far we have considered only the signals in the $V+W-2Q$ combination maps. In order to assess how much of this signal is affecting the CMB a first order approximation would be to take the equivalent of the internal linear combination but for the signal considered at the three WMAP frequencies. The best CMB map (or internal linear combination) was extracted by the WMAP team by combining the different bands with some optimal weights that reduce the foregrounds while keeping the CMB signal (see for instance Eriksen et al. (2004) where the authors compare the weights with two different methods and show how the internal linear combination is mostly dominated by the data in the V band). For our purposes we will assume the weights derived by Eriksen et al. (2004). Our internal linear combination will be then $1.925\times V - 0.468\times Q - 0.383\times W$. By looking at the quadrupole and octupoles of the simulated ZLE plus US (with fluxes less than 0.5 Jy) we find that the quadrupole has very little power (amplitude less than 1 $\mu K$ or about 3\% of the measured amplitude from WMAP data). A similar situation is observed when we compare the measured octupole and the octupole due to the contribution from the ZLE and US. Regarding the possible contribution from the Sunyaev-Zel'dovich effect in the Virgo region, as noted by Dolag et al. (2005), the contribution from Virgo (and other nearby structures) is far too small to affect significantly the measurements of the quadrupole and octupole.

\begin{figure}  
   \epsfysize=12cm   
   \begin{minipage}{\epsfysize}\epsffile{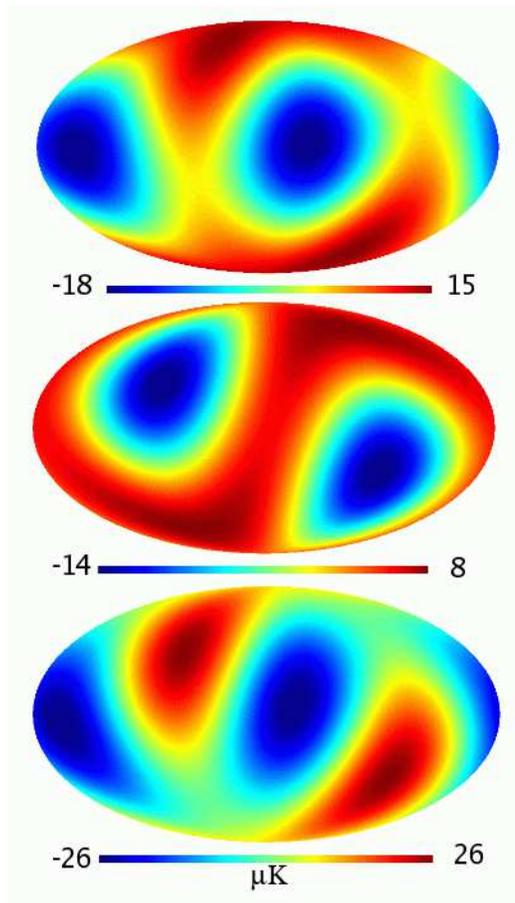}\end{minipage}  
   \caption{From top to bottom, WMAP 5yr quadrupole $(Q_{WMAP})$, quadrupole 
            of the hypothetical signal in the ecliptic plane $(Q_{ecl})$ 
            and the difference $Q_{WMAP}-Q_{ecl}$. 
           }  
   \label{fig_Quadrupoles}  
\end{figure}

Even though our models can not explain the excess in $V+W-2Q$, the excess is real and it must be caused by some real signal. As a toy model, it is interesting to speculate about the possibility that such a signal is proportional to the ZLE (or another signal concentrated around the ecliptic plane). If so, this could explain the low amplitude of the quadrupole and maybe the unusual alignment with the octupole. To test this we assume a hypothetical signal with a spatial distribution similar to the ZLE (see figure \ref{fig_Zodiacal}). We model the frequency dependence of this signal as $\nu ^{2.1}$, that is, very similar to a black body and choose its normalization such that its amplitude is approximately  10 $\mu K$ in $V+W-2Q$ band (after filtering with a $7^{\circ}$ Gaussian). This signal would explain the magnitude of the redder zones observed by WMAP. Since the signal behaves almost like a blackbody it would be difficult to detect in the component separation process and it would be confused as a CMB signal. With this toy model we compute the contribution to the internal linear combination using the same weights as above ($1.925\times V - 0.468\times Q - 0.383\times W$). The final map has a maximum amplitude of 31 $\mu K$ when filtered with a  $7^{\circ}$ Gaussian which is well below the maximum of the internal linear combination map for the CMB (range $\sim \pm 130$ $\mu K$) when filtered with the same scale. We can now derive the quadrupole and octupole of this map and compare with the observed values. We find that the quadrupole has significant power and is anti-correlated with the measured quadrupole in WMAP 5yr data (see figure \ref{fig_Quadrupoles}). This anti-correlation would explain why the quadrupole has an unusually low amplitude (Hinshaw et al. 1996, Bennett et al. 2003) as this hypothetical signal is suppressing a significant amount of the power. In figure \ref{fig_Quadrupoles} we show how the CMB quadrupole would look like if the contribution from this signal is subtracted from the measured quadrupole. We see how the amplitude of the CMB quadrupole increases nearly a factor two. Regarding the octupole we found that a signal like this contributes with a negligible amplitude ($\approx 1 \mu K$). We also found that only even multipoles have a significant amplitude and affect the measured power of the CMB while the effect from the odd multipoles (like the octupole) can be neglected. The anomalous signal, modifies slightly the direction of the quadrupole and could perhaps explain the anomalous quadrupole-octupole alignment. A careful study about this issue will be performed in a future paper.

\section{Discussion and Conclusions}  
Our analysis of the large scale residuals in the combination map $V+W-2Q$ reveals a signal which is inconsistent with the expected instrumental noise. This signal could be in principle explained with a variety of spectra ranging from blackbody-like spectra to dust-like spectra. It appears in areas with low Galactic contamination suggesting that is either from extragalactic origin or from our Solar System. Two possible explanations are explored to explain this signal;  Zodiacal light emission (ZLE) and unresolved compact sources (US). Neither the ZLE nor the US alone seem to be able to explain the magnitude of the observed high Galactic latitude signals. However, the models considered in this work are not very well constrained, hence significant variations in the expected intensity of the ZLE and US are still possible. We investigate alternative explanations like residual emission from the Galaxy and the Sunyaev-Zel'dovich effect but these possibilities are not able to explain in a convincing way the redder zones. However a combination of the effects is still possible in order to explain at least the redder zone near the Virgo region. The southern redder zone is harder to explain as there is no super-bright cluster (like Virgo) in this region.  This issue will need to be resolved with future data, in particular with Planck data which will cover a wider range of frequencies.     
Finally, we investigate the possible impact of the non-removed ZLE and US in the determination of the quadrupole and octupole in WMAP data. Our results show that the measurements of the quadrupole and octupole should not be affected by the non-removed ZLE or US. However, an alternative model with a quasi-blackbody spectrum and with a spatial distribution similar to the ZLE could explain both the redder zones and the low-l anomalies previously detected in the WMAP data. 

\section{Acknowledgments}  
JMD benefits from a Ministerio de Educaci\'on y Ciencia "Ram\'on y Cajal" 
contract. JMD and MC also acknowledge support from the Ministerio de Educaci\'on y Ciencia project AYA2007-68058-C03-02. Partial financial support for this research has been provided to JGN, MM and CB by the Italian ASI (contracts Planck LFI Activity of Phase E2 and de ASI contract I/016/07/0 COFIS) and MUR. Some of the results in this paper have been derived using the HEALPix (Gorski et al., 2005) package

  


\bsp  
\label{lastpage}  
\end{document}